\magnification=1200
\baselineskip=20 pt

\def\lamto{\lambda_{211}}
\def\lamth{\lambda_{311}}
\def\tilel{{\tilde e}_L}
\def\tiler{{\tilde e}_R}
\def\tilb{\tilde B}
\def\tilw{{\tilde W}_3}
\def\gm{\gamma}
\def\gpr{g^{\prime}}
\def\numul{\nu_{\mu L}}
\def\numulb{{\bar \nu}_{\mu L}}
\def\nutaul{\nu_{\tau L}}
\def\nutaulb{{\bar \nu}_{\tau L}}
\def\numu{\nu_{\mu}}
\def\nutau{\nu_{\tau}}
\def\lam{\lambda}
\centerline{\bf The effect of R parity violating couplings on 
$e^-e^-\rightarrow {\tilde e}_L {\tilde e}_R$}

\vskip 1 true in

\centerline{\bf Uma Mahanta}

\centerline{\bf Mehta Research Institute}

\centerline{\bf Chhatnag Road, Jhusi}

\centerline{\bf Allahabad-211019, India}

\vskip .4true in

\centerline{\bf Abstract}

In this work we study the effect of R parity breaking couplings on the
fermion number violating process $e^-e^-\rightarrow {\tilde e}_L 
{\tilde e}_R$. We find that an $e^-e^-$ linear collider operating at
$\sqrt {s}=500 $ Gev with an integrated luminosity of 50 fb$^{-1}$
will be able to probe the coupling $(\lamto ^2+\lamth ^3)^{1\over 2}$
down to .045 for a bino mass of 100 Gev and selectron mass of 200 Gev.
This would improve the present bound on it by a factor of 3.5.
More improved bounds can be obtained from precision measurement
of $\sigma_{LR}-\sigma_{RR}$ which reduces the MSSM background.

\vfill\eject

The process  $e^-e^-\rightarrow {\tilde e}_L {\tilde e}_R$ violates 
fermion number by two units. In the context of MSSM [1] with conserved R
parity it occurs in the lowest order via the t channel exchange of
a bino which we shall assume to be the lightest neutralino in our
study. The fermion number violation occurs because the $\tilde B$
being a Majorana fermion contributes a fermion number violating propagator
to the transition matrix element. Suppose the MSSM Lagrangian is now
extended by the R parity violating term that involves the product of three
lepton superfields [2]. The process  $e^-e^-\rightarrow {\tilde e}_L 
{\tilde e}_R$ then receives an additional contribution from thr t channel
exchange of a $\nu_{\mu}$ or $\nu_{\tau}$ even if they are assumed to be
Dirac fermions having only fermion number conserving propagators. 
The fermion number
violation in this case arises from the R parity breaking term that
violates fermion number precisely by two units. 

Consider the part of the MSSM Lagrangian that gives the interaction of
$\tilde B$ and $\tilde W_3$ with the $e-\tilde e$ pair [1].

$$L_0={1\over \sqrt 2}[ {\bar e}_R(g\tilw +\gpr \tilb ) \tilel 
+2 \gpr   {\bar e}_L\tilb\tiler ] + h.c. \eqno(1)$$

Among $\tilde B$ and ${\tilde W}_3$, only  $\tilde B$ interacts both with
$\tilel$ and $\tiler$. In the context of MSSM the process 
$e^-e^-\rightarrow {\tilde e}_L {\tilde e}_R$ can therefore occur through
the t channel exchange of $\tilde B$ only. The cross section for the process
$e^-e^-\rightarrow {\tilde e}_L {\tilde e}_R$ was first computed by Keung and
Littenburg [3]. The matrix elements for t channel and u channel exchange of
$\tilde B$ are given by 

$$M_a=g^{\prime 2}{\bar v}(p_2, s_2){(p_1-k_1).\gm\over (t-M^2)}P_L
u(p_1, s_1).\eqno(2a)$$

and

$$M_b=g^{\prime 2}{\bar v}(p_2, s_2){(p_1-k_2).\gm\over (u-M^2)}P_R
u(p_1, s_1).\eqno(2b)$$.

In the above $(p_1, s_1)$ and  $(p_2, s_2)$ are the momenta and spins of the
two incoming electrons. $k_1$ and $k_2$ are the momenta of the outgoing
 $\tilel$ and $\tiler$ respectively.
M is the mass of $\tilb$. $t=(p_1-k_1)^2=(p_2-k_2)^2$
and $u=(p_1-k_2)^2=(p_2-k_1)^2$. $(p_1-k_1).\gm =(p_1-k_1)^{\mu}\gm_{\mu}$
and  $(p_1-k_2).\gm =(p_1-k_2)^{\mu}\gm_{\mu}$.
 To simplify our analysis we shall assume 
that the mass of $\tilel$ and $\tiler$  are equal and we shall denote their
common mass by m.

If the incoming electron beams are both unpolarized then  the total matrix
 element for the process is given by $M=M_a+M_b$. The differential scattering
cross-section for the MSSM contribution is given by

$${d\sigma_0\over dy}={1\over 128\pi s}g^{\prime 4} (1-a-y^2)
[{1\over (1-b-y)^2}+{1\over (1-b+y)^2}].\eqno(3)$$

Here $y=\beta x =(1-4{m^2\over s})^{1\over 2} \cos\theta$, $a={m^2\over s}$
and $b={m^2-M^2\over s}$. $\sigma_0$ is the MSSM contribution to the 
cross section.
Note that in the MSSM contribution to $e^-e^-\rightarrow\tilel\tiler$
there is no interference between t channel and u channel exchanges.

Let us now consider the MSSM Lagrangian to be extended by the following R
parity violating terms

$$L_1=[\lamto \{\tilel{\bar e}_R\numul   +\tiler ^*(\numulb )^c e_L \}
  +\lamth   \{\tilel{\bar e}_R\nutaul +\tiler ^* (\nutaulb )^c e_L \} ]
+h.c. \eqno(4)$$

The process $e^-e^-\rightarrow \tilel\tiler$ violates fermion number by
two units but it conserves R parity . The transition matrix element should 
therefore involve an even number of R parity violating interaction vertices.
In the above Lagrangian the term $\lamto \tiler ^* (\numulb )^c e_L$ violates
fermion number by two units but the term $\lamto ^*\tilel^* \numulb e_R$
conserves fermion number. Hence the product of 
$\lamto \tiler ^* (\numulb )^c e_L$ and $\lamto ^*\tilel^* \numulb e_R$
which occurs in second order perturbation expansion can contribute to
$e^-e^-\rightarrow \tilel\tiler$. This also holds true for analogous
terms proportional to $\lamth$ and involving $\nu_{\tau}$.
Note that $ (\lamto \tiler^* (\numulb )^c e_L )^2$  or
 $ (\lamth \tiler^* (\nutaulb )^c e_L )^2$ which also occur in second order
pertubation expansion cannot contribute to the process 
$e^-e^-\rightarrow \tiler\tiler$ if the neutrinos are assumed to be Dirac
fermions having only fermion number conserving propagators. Similarly
$(\lamto \tilel^*\numulb e_R)^2$ or $(\lamth \tilel^*\nutaulb e_R)^2$
cannot contribute to the process $e^-e^-\rightarrow \tilel\tilel$.
Hence R parity violating couplings affect only
 the process $e^-e^-\rightarrow \tilel\tiler$ leaving 
 $e^-e^-\rightarrow \tilel\tilel$ and $e^-e^-\rightarrow \tiler\tiler$
unchanged.

We shall now evaluate the transition matrix elements arising from R parity 
violating couplings. Consider first the t channel exchange of a $\nu_{\mu}$.
We have

$$\eqalignno{-i\delta M_a^{\mu} &=-\vert\lamto\vert^2 <\tiler (k_2)\tilel (k_1)
\vert \tiler^*(\numulb )^c e_L \tilel^* \numulb e_R 
\vert e(p_1,s_1);e(p_2,s_2)>\cr
&=\vert\lamto\vert^2<0\vert \nu_{\mu}^T C^+P_Le{\bar\nu}_{\mu}P_Re\vert
e(p_1,s_1)e(p_2,s_2)>\cr
&=-\vert\lamto\vert^2 u^T(p_2,s_2)P_L^TC^{+T}<0\vert T \{ \nu_{\mu}
{\bar\nu}_{\mu}\}\vert 0>P_R u(p_1,s_1)\cr
&=\vert\lamto\vert^2 {\bar v}(p_2,s_2)CP_L^TC^{+T}{i(p_1-k_1).\gm\over
(p_1-k_1)^2}P_R u(p_1,s_1)\cr
&=-i\vert\lamto\vert^2{\bar v}(p_2,s_2){(p_1-k_1).\gm\over t}P_R 
u(p_1,s_1).&(5)\cr}$$

Adding the contribution due to t channel exchange of $\nu_{\tau}$
we get

$$\delta M_a={\vert\lamto\vert^2+\vert\lamth\vert^2\over t}{\bar v}(p_2,s_2)
(p_1-k_1).\gm P_R u(p_1,s_1).\eqno(6)$$

Consider next the matrix element due to u channel exchange of $\numu$. We
have

$$\eqalignno{-i\delta M_b^{\mu}&=-\vert\lamto
     \vert^2<\tiler (k_2)\tilel (k_1)\vert
\tilel^* \numulb e_R \tiler^*(\numulb )^c e_L\vert e(p_1,s_1);e(p_2,s_2)>\cr
&=\vert\lamto\vert^2 <0\vert {\bar\nu}_{\mu}P_Re \numu^T C^+P_L e
\vert e(p_1,s_1);e(p_2,s_2)>\cr
&=-\vert\lamto\vert^2 u^T(p_2,s_2)P_R^T<0\vert T \{ {\bar\nu}_{\mu}^T
\nu_{\mu}^T\}\vert 0>C^+P_Lu(p_1,s_1)\cr
&=\vert\lamto\vert^2 u^T(p_2,s_2)P_R^T<0\vert T\{ \numu{\bar\nu}_{\mu}\}^T
\vert 0>C^+P_L u(p_1,s_1)\cr
&=-\vert\lamto\vert^2{\bar v}(p_2,s_2)CP_R^T{i(p_2-k_1).\gm\over u}^TC^+
P_Lu(p_1,s_1)\cr
&=-i{\vert\lamto\vert^2\over u} {\bar v}(p_2,s_2) (p_1-k_2).\gm P_L 
u(p_1,s_1).&(7)\cr}$$

Adding the contribution of $\nutau$ we get

$$\delta M_b= {\vert\lamto\vert^2+\vert\lamth\vert^2\over u}
{\bar v}(p_2,s_2)
(p_1-k_2).\gm P_L u(p_1,s_1).\eqno(8)$$

Therefore the total matrix element arising from R parity violating 
couplings is given by
 
$$\delta M = (\vert\lamto\vert^2+\vert\lamth\vert^2){\bar v}(p_2,s_2)
[{(p_1-k_1).\gm\over t} P_R+{(p_1-k_2).\gm\over u} P_L]u(p_1,s_1).
\eqno(9)$$

In deriving the matrix elements due to R parity violating couplings
 particular care has been taken in transposing
a pair of Fermi fields because of their anticommuting nature.

The differential scattering cross section due to MMSM plus R parity 
violating couplings is given by 

$$\eqalignno{ {d\sigma_t\over dy}&={g^{\prime 4}\over 128\pi s}[ (1-a-y^2)
\{{1\over (1-b-y)^2}+{1\over (1-b+y)^2}\}\cr
&+{\lambda^4\over g^{\prime 4}}(1-a-y^2)\{ {1\over (1-a-y)^2}+{1\over 
(1-a+y)^2} \}\cr
&+{\lambda^2\over g^{\prime 2}}(1+2a+2y^2)\{ {1\over (1-a+y)(1-b-y)}+
{1\over (1-a-y)(1-b+y)}\}].&(10)\cr}$$

where $\lambda^2 =[\lamto^2+\lamth^2]$.

In MSSM the lightest neutralino usually also happens to be the lightest 
supersymmetric particle (LSP). For a bino mass of 100 Gev, selectron
mass of 200 Gev and ${\sqrt s}$=500 Gev we find that $\sigma_0$=240.82 fb
and $\sigma_t =\sigma_0+\delta\sigma = (240.82+3117.8\lambda^2+
16998.7\lambda^4)$ fb. A bound on $\lam$ can be derived by requiring that for
observability, the signal $\delta \sigma \int L dt$ should be
at least greater than or equal to $3[ \sigma_0\int L dt]^{1\over 2}$. For an 
$e^-e^-$ linear collider operating at ${\sqrt s}$=500 Gev with an integrated
luminosity of of 50 fb$^{-1}$ this produces the bound $\lam \le$ .045. 
The present best bound on $\lam$ for m=200 Gev is .156 [4]. The bound 
obtainable from an $e^-e^-$ collider under the above conditions will therefore
improve the present bound by a factor of 3.5. The present bounds on
R parity breaking couplings which are mostly obtained from low energy
measurements scale with m. Hence by choosing higher values of $\sqrt s$
which would enable the production of heavier selectrons, the collider
bound can be much smaller than the present bound. 
 In some supersymmetric models
as for example in scenarios with gauge mediated supersymmetry breaking
the selectron often turns out to be the LSP. For a selectron mass of 150
Gev, bino mass of 250 Gev and ${\sqrt s}$ =500 Gev we find that 
$\sigma_0$=173.69 fb and $\delta\sigma =(3718.72\lam^2+31478.2\lam^4)$ fb. 
Under the same operating conditions as above ( ${\sqrt s }$=500 Gev
and $\int L dt$= 50 fb$^{-1}$ we obtain the bound $\lam \le .039$. This 
should be compared with the present bound of .117 for m=150 Gev.
More improved bound on $\lam$ can be derived from precision measurement
of $\sigma_{LR}-\sigma_{RR}$ where $\sigma_{LR}$ is the cross section
for $e^-e^-\rightarrow \tilel\tiler$ and $\sigma_{RR}$is that of
$e^-e^-\rightarrow \tilel\tiler$. Since R parity violating couplings
affect only $\sigma_{LR}$ the MSSM background for $\sigma_{LR}-\sigma_{RR}$
will be lower than that for $\sigma_{LR}$. The reduction of MSSM background
will however be significant only if the bino mass is not too small compared
to $\sqrt s$. The reason being $\sigma_{RR}$ receives contribution from
the chirality flipping part of the bino propagator and is proportional
to ${M^2\over s}$. For m=150 Gev, M=250 G
ev and $\sqrt s$= 500 Gev
 we find that $\sigma_{RR}$
=241.01 fb. Hence under the same operating conditions as before,
 from precision 
measurement of $\sigma_{LR}-\sigma_{RR}$ it should be possible to probe
$\lam $ down to .030.

In conclusion in this work we have studied the effect of R parity breaking
couplings on the fermion number violating process $e^-e^-\rightarrow\tilel
\tiler$. We find that an $e^-e^-$ linear collider operating at $\sqrt s$=
500 Gev with an integrated luminosity of 50 fb$^{-1}$ should be able to
probe the coupling $[\lamto^2+\lamth^2]^{1\over 2}$ down to .045 for M=100
Gev and m=200 Gev. This bound can be obtained simply from precision 
measurement of the total cross section $\sigma_{LR}$.
More improved bounds can be derived from precision
measurement of $\sigma_{LR}-\sigma_{RR}$ which reduces the MSSM background.
The $e^-e^-$ mode of a 
linear collider has recently been proposed [5] for high
precision measurement of superparticle couplings and verfying the
 supersymmetric relation between ordinary couplings and superparticle
couplings. Our study clearly shows that to measure bino couplings
precisely one should use the process $e^-e^-\rightarrow \tiler\tiler$
instead of $e^-e^-\rightarrow \tilel\tiler$
since the former as opposed to the later is not affected even if R 
parity violating couplings are present.

\centerline{\bf References}

\item{1.} H. P. Nilles, Phys. Rep. 110, 1 (1984); H. E. Haber and 
G. L. Kane, Phys. Rep. 117, 75, (1985).

\item{2.} F. Zwirner, Phys. Lett. B 132, 103 (1983); L. J. Hall and
M. Suzuki, Nucl. Phys. B 231, 419 (1984); G. G. Ross and J. W. F. Valle,
Phys. Lett. B 151, 375 (1985); S. Dawson, Nucl. Phys. B 261, 297 (1985);
V. Barger, G. F. Giudice and T. Han, Phys. Rev. D 40, 2987 (1989).

\item{3.} W. Y. Keung and L. Littenburg, Phys. Rev. D 28, 1067 (1983);
H. E. Haber and G. L. Kane, Phys. Rep. 117, 75 (1985).

\item{4.} H. Dreiner, hep-ph/9707435.

\item{5.} H. C. Cheng, J. L. Feng and N. Polonsky, Phys. Rev. D 57, 152 
(1998); H. C. Cheng, hep-ph/9801234.

\end